\documentclass[preprint2]{aastex}

\newcommand{\ergcms}{ergs cm$^{-2}$ s$^{-1}$\ }
\newcommand{\ergcmsnospace}{ergs cm$^{-2}$ s$^{-1}$}

\newcommand{\arielv}{{\em Ariel V}}
\newcommand{\asca}{{\em ASCA\ }}
\newcommand{\sax}{{\em BeppoSAX\ }}
\newcommand{\saxnospace}{{\em BeppoSAX}}
\newcommand{\chandra}{{\em Chandra\ }}
\newcommand{\chandranospace}{{\em Chandra}}
\newcommand{\einstein}{{\em Einstein\ }}
\newcommand{\exosat}{{\em EXOSAT\ }}
\newcommand{\granat}{{\em Granat\ }}
\newcommand{\heao}{{\em HEAO-1}}
\newcommand{\mir}{{\em MIR}}
\newcommand{\oso}{{\em OSO-7}}
\newcommand{\rosat}{{\em ROSAT\ }}
\newcommand{\rxte}{{\em RXTE\ }}
\newcommand{\uhuru}{{\em Uhuru}}
\newcommand{\sas}{{\em SAS-3\ }}

\begin{document}
\title{\chandra Observations of the Faintest Low-Mass X-ray Binaries}
\author{Colleen. A. Wilson\altaffilmark{1}, Sandeep K. Patel\altaffilmark{2},
Chryssa Kouveliotou\altaffilmark{1,3}}
\affil{SD 50 Space Science Research Center, National Space Science and 
Technology Center, 320 Sparkman Drive, Huntsville, AL 35805}
\email{colleen.wilson-hodge@nsstc.nasa.gov}
\author{Peter G. Jonker}
\affil{Institute of Astronomy, Cambridge University, Madingley Road, CB30HA,
Cambridge, United Kingdom}
\author{Michiel van der Klis}
\affil{Astronomical Institute 'Anton Pannekoek', University of Amsterdam and Centre for High-Energy Astrophysics, Kruislaan 403, 1098 SJ
Amsterdam, the Netherlands}
\author{Walter H.G Lewin}
\affil{Department of Physics and Center for Space Research, Massachusetts
Institute of Technology, Cambridge, MA 02138}
\author{Tomaso Belloni}
\affil{Osservatorio Astronomico di Brera, Via E. Bianchi 46, 23807 Merate
(LC), Italy} 
\author{Mariano Mendez}
\affil{SRON National Institute for Space Research, Sorbonnelaan 2, 3584 CA
Utrecht, The Netherlands}
\altaffiltext{1}{NASA's Marshall Space Flight Center}
\altaffiltext{2}{National Academy of Sciences NRC Fellow}
\altaffiltext{3}{Universities Space Research Association}
\begin{abstract}
There exists a group of persistently faint galactic X-ray sources that, based on
their location in the galaxy, high $L_{\rm x}/L_{\rm opt}$, association with X-ray bursts, and 
absence of low frequency X-ray pulsations, are thought to be low-mass X-ray binaries (LMXBs). 
We present results from Chandra observations for eight of these systems: 4U 1708--408, 
2S 1711--339, KS 1739--304, SLX 1735--269, GRS 1736--297, SLX 1746--331, 1E 
1746.7--3224, and 4U 1812--12. Locations for all sources, excluding GRS
1736--297, SLX 1746--331, and KS 1739--304 (which were not detected) were
improved to 0.6\arcsec\ error circles (90\% confidence). Our observations
support earlier findings of transient behavior of GRS 1736-297,
KS 1739--304, SLX 1746-331, and 2S 1711-339 (which we detect in one of two 
observations). Energy spectra for 4U 1708--408, 2S 1711--339, SLX 1735--269, 
1E 1746.7-3224, and 4U 1812--12 are hard, with power law indices typically
1.4-2.1, which are consistent with typical faint LMXB spectra. 
\end{abstract}

\section{Introduction}
Low mass X-ray binaries (LMXBs) are systems in which a low mass 
($<1$ M$_{\odot}$) star transfers matter to either a low magnetic field 
(typically $\sim 10^9$ G) neutron star or a black hole. In both cases plasma 
flows down to a few stellar radii and produces observable properties in X-rays 
(spectra and timing). Over the last several years comprehensive studies of these
properties have provided crucial information on fundamental properties of 
compact objects, e.g., the equation of state of neutron stars (see van der Klis
2000\nocite{vanderKlis00} for a review), or general relativistic effects near black hole horizons 
(see Tanaka \& Lewin 1995,\nocite{Tanaka95} Esin et al. 2001\nocite{Esin01}). 
However, there exist several sources whose X-ray properties indicate they may 
also be LMXBs but whose X-ray emission was so faint that no instruments were 
sensitive enough to study them until the launch of \chandranospace. This paper 
describes our \chandra observations of eight such objects listed as candidate 
LMXBs in \citet{vanParadijs95}; the accurate locations we derived with \chandra will 
enable follow up observations in the infrared/optical. 

Below we briefly summarize prior observations for each of these objects. 
Table~\ref{tab:fluxsum} provides a convenient comparison of the observed fluxes
discussed below with our \chandra observations. 

\paragraph{4U 1708--40} is a persistent X-ray source that was observed by most
of the early X-ray instruments, e.g. \uhuru, \oso, \arielv, and \heao\
\citep[and references therein]{vanParadijs95}. A \rosat source, 1RXS J17141224.8--405034,
at R.A. $17^h12^m24^s.8$, Decl. = --40\arcdeg50\arcmin34\arcsec.5 (J2000,
1-$\sigma$ error radius = 10\arcsec), detected in the \rosat all-sky survey, lies
in the error box for 4U 1708--40 \citep{Voges99}. Since 1996, 4U 1708--40
has been a persistent source in the {\em Rossi X-ray Timing Explorer (RXTE)}
All-Sky Monitor (ASM). Its 2-10 keV flux\footnote{Throughout this 
paper we use the conversion factors 1 mCrab = 0.75 ASM ct/s = $2.08 \times 
10^{-11}$ \ergcms (2-10 keV).} slowly declined from about $6.2 \times 10^{-10}$
\ergcms in 1996-1997 to a minimum of about $1.0 \times 10^{-10}$ \ergcms in 
mid-1998, followed by a slow rise to about $8.3 \times 10^{-10}$ \ergcms in 
mid-2001, and a second slow decline that continued in 2002-2003. At the time of 
our \chandra observations, the 2-10 keV flux level in the \rxte ASM was about 
$5.2 \times 10^{-10}$ \ergcmsnospace. \citet{Migliari03} reported the discovery of 
X-ray bursts in a \sax observation of 4U 1708--408 in
1999 August. During this observation, the spectrum of the persistent flux
(before and after the bursts) was well fitted with an absorbed power-law with
a column density $N_{\rm H} = (2.93 \pm 0.08) \times 10^{22}$ cm$^{-2}$,
a photon index of $2.42 \pm 0.02$, and a Gaussian iron line. The unabsorbed 2-10
keV flux was $1.2 \times 10^{-10}$ \ergcmsnospace. In addition, 
\citet{Migliari03} also reported results for several \rxte observations in 1997
and 2000 along with an additional \sax observation in 2001 during which no 
bursts were seen; however, steady persistent emission was observed. The energy 
spectra in these observations required more complicated models, an absorbed 
power law with photon index $\sim 2.7$ for an assumed $N_{\rm H} = 2.9 \times 
10^{22}$ cm$^{-2}$, and a blackbody component with  $kT \sim 1.3$ keV. The 
unabsorbed 2-10 keV fluxes were $7.4 \times 10^{-10}$ \ergcms and $8.9 \times 
10^{-10}$ \ergcms in the 2000 June 18 \rxte and 2001 August \sax observations, 
respectively.

\paragraph{2S 1711--339} is an X-ray burster originally detected with \arielv\
\citep{Carpenter77} and more precisely located with \sas \citep{Greenhill79}.
A radio source was detected in the \sas error circle at R.A. = $17^h10^m52^s$,
Decl. = --34\arcdeg00\arcmin36\arcsec (B1950, 90\% confidence error radius 
2.2\arcmin) in 1978 July \citep{Greenhill79}. A \rosat source, RXS
J171419.3--34023 \citep{Voges99}, also lies within the \sas error circle, but 
just outside the radio source error circle. More
recently, \citet{Cornelisse02} reported on observations of 2S 1711--339 with
\saxnospace, the \rxte ASM, and {\em Chandra}. The \rxte ASM detected an X-ray outburst
of 2S 1711--339 from 1998 July until 1999 May at a 2-10 keV flux level of $8.2
\times 10^{-10}$ \ergcmsnospace. During that time period, \sax 
detected 10 short X-ray bursts, with a persistent flux level (before and after 
the bursts) of $6.3 \times 10^{-10}$ \ergcms (2-28 keV). The persistent emission
was fitted with a cutoff power-law with photon index 0.7, high energy cutoff 2.8
keV, and an assumed $N_H$ of $1.5 \times 10^{22}$ cm$^{-2}$, 
derived from \sax Narrow Field Instrument (NFI) observations on 2000 February 
29, when the 2-6 keV flux was $2.4 \times 10^{-11}$ \ergcmsnospace. On 2000 March 22, one burst was detected with \saxnospace, but the 3-$\sigma$ upper
limit on the persistent flux (before and after the burst) was $\lesssim 7 \times
10^{-11}$ \ergcms (2-28 keV). The \chandra observation, also described in this 
paper, on 2000 June 9, yielded a location of R.A. = $17^h14^m19.8^s$, Decl. = 
--34\arcdeg02\arcmin47\arcsec\ (J2000, 90\% confidence error radius 1\arcsec), 
which was consistent with the Wide Field Camera \citep{Cornelisse02}, \arielv\ 
\citep{Carpenter77}, and \rosat \citep{Voges99} positions. The \chandra
position, however, lies 0.2\arcmin\ outside the 90\% confidence error circle of
the radio source.  

\paragraph{SLX 1735--269} is an X-ray burster first reported in 1985 from
Spacelab-2 X-ray Telescope (XRT) observations \citep{Skinner87}. It was present in the
\einstein slew survey 5 years earlier \citep{Elvis92}. A \rosat source, 1RXS
J173817.0--265940, lies within the \einstein error circle \citep{Voges99}.
\granat Sigma observations showed it to be a persistent hard X-ray source
\citep{Goldwurm96}. Type I X-ray bursts were discovered with the \sax
Wide Field Camera \citep[WFC]{Bazzano97}. \asca observations showed the 0.6-10 keV spectrum to be
well fitted with a power-law with photon index 2.15 and $N_{\rm H} \sim
(1.4-1.5) \times 10^{22}$ cm$^{-2}$ \citep{David97}, which was consistent with
the values derived from \rosat and
\granat ART-P
observations \citep{Pavilinsky94, Grebenev96}. Using \rxte observations from 1997 February-May, 
\citet{Wijnands99a} showed that the power
spectrum was characterized by a strong band limited noise component which was
approximately flat below a break frequency of 0.1-2.3 Hz. Above the break
frequency the power spectrum declines as a power law of index 0.9. At the
highest count rates, a broad bump was observed around 0.9 Hz. The strength of
the noise (2-60 keV, integrated over 0.01-100 Hz) was $\sim 24$\% and 
$\sim 17$\% rms in the high and low count rate data, respectively. Fits to energy 
spectra yielded power law indices of 2.2 and 2.4 (with an assumed $N_H$ of 
$1.47 \times 10^{22}$ cm$^{-2}$) and 3-25 keV fluxes of $3.8 \times 10^{-10}$ 
\ergcms and $2.8 \times 10^{-10}$ \ergcmsnospace, respectively. Power spectra, from 
extended \rxte observations in 1997 October \citep{Barret00} at a 1-20 keV 
flux level of $\sim 7.2 \times 10^{-10}$ \ergcmsnospace, had a noise strength of
27.6\% (2-40 keV, integrated over 0.005-300 Hz) and a break frequency of
0.08-0.15 Hz that appeared to increase with intensity. \citet{Belloni02} fitted
this power spectrum with a four-Lorentzian model and found a remarkable
resemblance to observations of the low-luminosity X-ray burster 1E 1724--3045. 
SLX 1735--269 has been detected as a persistent source with the \rxte ASM since
1996 at an average 2-10 keV flux level of about $3.1 \times 10^{-10}$ 
\ergcmsnospace. 

\paragraph{GRS 1736--297} was discovered with the \granat ART-P telescope in
September 1990 and was subsequently observed about a month later
\citep{Pavilinsky92} at a similar flux level. Its spectrum was characterized by
a power law with photon index of 1.8 and a 3-12 keV flux of 
$6 \times 10^{-11}$ \ergcmsnospace. \citet{Motch98} identified RX J1739.5--2942 with
GRS 1736--297, since it lies well within the 90\% confidence 90\arcsec\ radius
\granat error circle \citep{Pavilinsky94}. Further, \citet{Motch98} identified a
Be star in the \rosat error circle, suggesting that, although GRS 1736--297 has
been previously classified as a candidate LMXB \citep{vanParadijs95}, it may 
instead be a transient Be/X-ray binary system.

\paragraph{KS 1739--304} is a transient X-ray source discovered with \mir/KVANT
in 1989 at a 2-30 keV flux level of about 9 mCrab and located to within a
1.6\arcmin\ error circle \citep{Sunyaev91, Cherepashchuk94}. 
Subsequent observations with \granat ART-P in Autumn 1990 did not detect the
source, with a 3-$\sigma$ upper limit of 2 mCrab (8-20 keV)
\citep{Pavilinsky94}. Very little is known about this object.

\paragraph{SLX 1746--331} was discovered with the Spacelab-2 XRT in 
1985 August \citep{Skinner90}. It had a very soft spectrum, best fitted with a 
thermal Bremsstrahlung with $kT = 1.5$ keV and a 2-10 keV flux of $6.35 \times 
10^{-10}$ \ergcmsnospace. Based on the very soft spectrum, \citet{Skinner90} 
suggested SLX 1746--331 as a potential black hole candidate. The position of 
SLX 1746--331 fell between fields in the \einstein galactic plane survey of 
\citet{Hertz84} and it was not detected in the \exosat galactic plane survey 
\citep{Warwick88}. During the \rosat
survey observation in 1990 September, 1RXS J174948.4--331215, 
which lies in the error circle of SLX 1746--331, was detected in
outburst, but it was not detected in subsequent \rosat High Resolution Imager 
(HRI) observations in 1994 October \citep{Motch98}. Hence the \exosat and \rosat
non-detections suggest this object is a transient system.

\paragraph{1E 1746.7--3224} was discovered during the \einstein galactic plane
survey \citep{Hertz84}, at a 2-10 keV flux level of $3 \times 10^{-12}$
\ergcmsnospace.
Spacelab-2 XRT observations in 1985 August gave an upper limit on the
2-10 keV flux of $\lesssim 4 \times 10^{-11}$ \ergcms \citep{Skinner90}. During
the \rosat all-sky survey, 1RXS J175003.8-322622 was detected in the \einstein 
error circle of 1E 1746.7--3224 \citep{Voges99}. Very little is known about
this object. 

\paragraph{4U 1812--12} was first detected with \uhuru\
\citep{Forman76, Forman78} and was later observed by several satellites,
including \oso, \arielv, \heao, and \exosat with a typical 2-10 keV flux of
$\sim 4 \times 10^{-10}$ \ergcms 
\citep[and references therein]{vanParadijs95}. 4U 1812--12 was detected in the 
\rosat All-Sky Survey as a bright source denoted 1RXS J181506.1--120545
\citep{Voges99}. X-ray bursts were first
detected from this source in 1982 with Hakucho \citep{Murakami83} and later with
\sax \citep{Cocchi00}. 4U 1812--12 is classified as an atoll source and
shows band limited noise with a break frequency of 0.095 Hz and a QPO at 0.85 Hz
\citep{Wijnands99b}. 4U 1812--12 has been
detected as a persistent source with the \rxte ASM since 1996 at an average
2-10 keV flux level of about $3.7 \times 10^{-10}$ \ergcmsnospace. Recent observations in 2000 April with 
\rxte and \sax show 4U 1812--12 in a hard state with a hard tail extending
above 100 keV. The power spectrum was characterized by a $\sim 0.7$ Hz QPO and 
three broad noise components, extending above $\sim 200$ Hz \citep{Barret03}.

\section{Observations and Results}

Each of the eight objects was observed using the Advanced CCD Imaging Spectrometer
(ACIS) on the \chandra X-ray observatory. Table~\ref{tab:obs} lists the details
of the observations. For some sources, data were collected in two different 
observing modes, timed exposure (TE) mode and continuous clocking (CC) mode. 
Data obtained in TE mode allow for two-dimensional imaging. Accurate spectroscopy
of bright targets, however, is limited due to pulse pile-up. Timing studies are
limited by the total number of photons collected and the 3.24 s time resolution.
In CC mode, the amount of pile-up is negligible owing to the 2.85 ms time 
resolution, allowing for accurate spectroscopy and timing. Unfortunately only 
one of the sources, SLX 1735--269, was detected in CC mode.

The best known location for each object was positioned on the nominal target
position of ACIS-S3, a back illuminated CCD on the spectroscopic array (ACIS-S)
with good charge transfer efficiency and spectral resolution, with the exception
of the CC mode observation of SLX 1735--269 which fell on ACIS-S2, a front illuminated
CCD on the spectroscopic array (ACIS-S). Standard processing was performed by
the \chandra data center. The data were filtered to exclude events with grades
1, 5, 7, hot pixels, bad columns, and events on CCD node boundaries. Analysis in
this paper was done using \chandra Interactive Analysis of Observations 
(CIAO)\footnote{http://cxc.harvard.edu/ciao/}
version 2.2.1 and \chandra Calibration Database 
(CALDB)\footnote{http://cxc.harvard.edu/caldb/index.html} version 2.10.
For piled-up sources, the tool {\em acis\_detect\_afterglow} in the standard
processing pipeline can reject valid source photons in addition to afterglow
photons resulting from residual charge from cosmic ray events in the CCD pixels.
We examined the spatial distribution of afterglow events for each observation 
using methods described in the CIAO documentation\footnote{http://asc.harvard.edu/ciao/threads/acisdetectafterglow/}.
For TE mode observations of 2S 1711--339 and 1E 1746.7--3224 we found that
$\gtrsim 5$\% of events consistent with the point source had been rejected; 
hence we reprocessed the data to retain the afterglow flagged events. For all 
other observations, $< 1$\% of events were rejected, so the standard processing
was retained.

\subsection{Source Locations}

Using TE mode data, we extracted source locations for each of the detected
sources. These locations are listed in Table~\ref{tab:loc}. Sources were
located by one of two methods: (1) 2S 1711-339 and 1E 1746.7--3224 were located
using the CIAO tool {\em wavdetect}. The latter source was observed twice in TE
mode and locations from both observations were consistent. (2) 4U 1708--40, 
SLX 1735--269, and 4U 1812--12 were brighter and suffered from considerable 
pile-up of photons in the image core, resulting in a source that looked ring 
shaped with a hole in the center. The CIAO Detect tools were inadequate to
provide a good location of the centroid, therefore we modeled the data using a 
Gaussian function multiplied by a hyperbolic tangent in radius, scaled to 
approach zero at r = 0.0 \citep{Hulleman01}.  The uncertainty of these locations
is limited by systematic effects to a circle with $\sim 0.6\arcsec\ $ radius 
\citep{Aldcroft00}. No other known sources were found in the images, hence we 
were unable to use astrometry to further improve the locations.

\subsection{Detection Upper Limits}

Three objects, GRS 1736--297, SLX 1746--331, and KS 1739--304 went undetected in
all observations. To derive upper limits for these sources, we used the CIAO 
tool {\em dmextract} to extract count rates for a 6\arcsec\ radius source 
region and a background annulus (inner radius = 6\arcsec, outer radius = 
30\arcsec) centered on the best known position.  The 99\% confidence upper limits 
were count rates of $4.9 \times 10^{-3}$ cts s$^{-1}$, $3.7 \times 10^{-3}$ cts 
s$^{-1}$, and $6.2 \times 10^{-3}$ cts s$^{-1}$ (total counts = 6.0 cts, 5.8
cts, and 4.6 cts) for GRS 1736--297, SLX 1746--331, and KS 1739--304, 
respectively.
Using WebPimms\footnote{http://heasarc.gsfc.nasa.gov/Tools/w3pimms.html}, we
estimated an upper limit 1-10 keV absorbed flux of $7.5 \times 10^{-14}$ \ergcms 
for GRS 1736--297, assuming $N_H = 1.1 \times 10^{22}$ cm$^{-2}$, derived using
 Colden\footnote{http://asc.harvard.edu/toolkit/colden.jsp}, a web tool based on
\citet{Dickey90}, and a power law
photon index = 1.8 \citep{Pavilinsky92}. Similarly, for SLX 1746--331, we
estimated an upper limit 1-10 keV absorbed flux of $2.0 \times 10^{-14}$
\ergcmsnospace, assuming $N_H = 0.4 \times 10^{22}$ cm$^{-2}$, using Colden, and a
thermal Bremsstrahlung spectrum with kT = 1.5 keV \citep{Skinner90}. For KS
1739--304, we estimated an upper limit 1-10 keV absorbed flux of $1.2 \times
10^{-13}$ \ergcmsnospace, assuming $N_H = 1.4 \times 10^{22}$ cm$^{-2}$, using Colden,
and a power law photon index = 1.5.  A fourth object, 2S 1711--339, that was 
detected in a TE mode observation in 2000, went undetected in the 2002 CC mode 
observation. To derive an upper limit for it, we again used dmextract, with a 
6\arcsec\ radius source region and a background annulus (inner radius = 
6\arcsec, outer radius = 30\arcsec) centered on the \chandra position (shifted 
to correspond to the 1-dimensional CC mode image) from the 2000 observation, to
obtain 99\% confidence upper limits of $6.5 \times 10^{-3}$ cts s$^{-1}$ or a total 
30.3 counts for the observation. Using WebPimms with our spectral parameters 
from the 2000 observation Table~\ref{tab:spec}, we estimate a 1-10 keV upper 
limit absorbed flux of $1 \times 10^{-13}$ \ergcmsnospace, a factor of more than 100 
fainter than in the 2000 observation. In the 5 May 2002 observation of KS
1739--304, an object was present at R.A. = $17^h42^m41^s.07$, Decl.=
$-30\arcdeg22\arcmin39.3\arcsec$, approximately $8.2\arcmin$\ from the expected
location of KS 1739--304. We measured a total of approximately 45 counts from
this object in the 745 s observation. Given its large separation from KS 
1739--304, and the fact that it is well outside the 1.6\arcmin\ radius
\mir/KVANT error circle \citep{Sunyaev91}, the two objects are most likely 
unrelated.

\subsection{Energy Spectra}

\begin{figure}
\plottwo{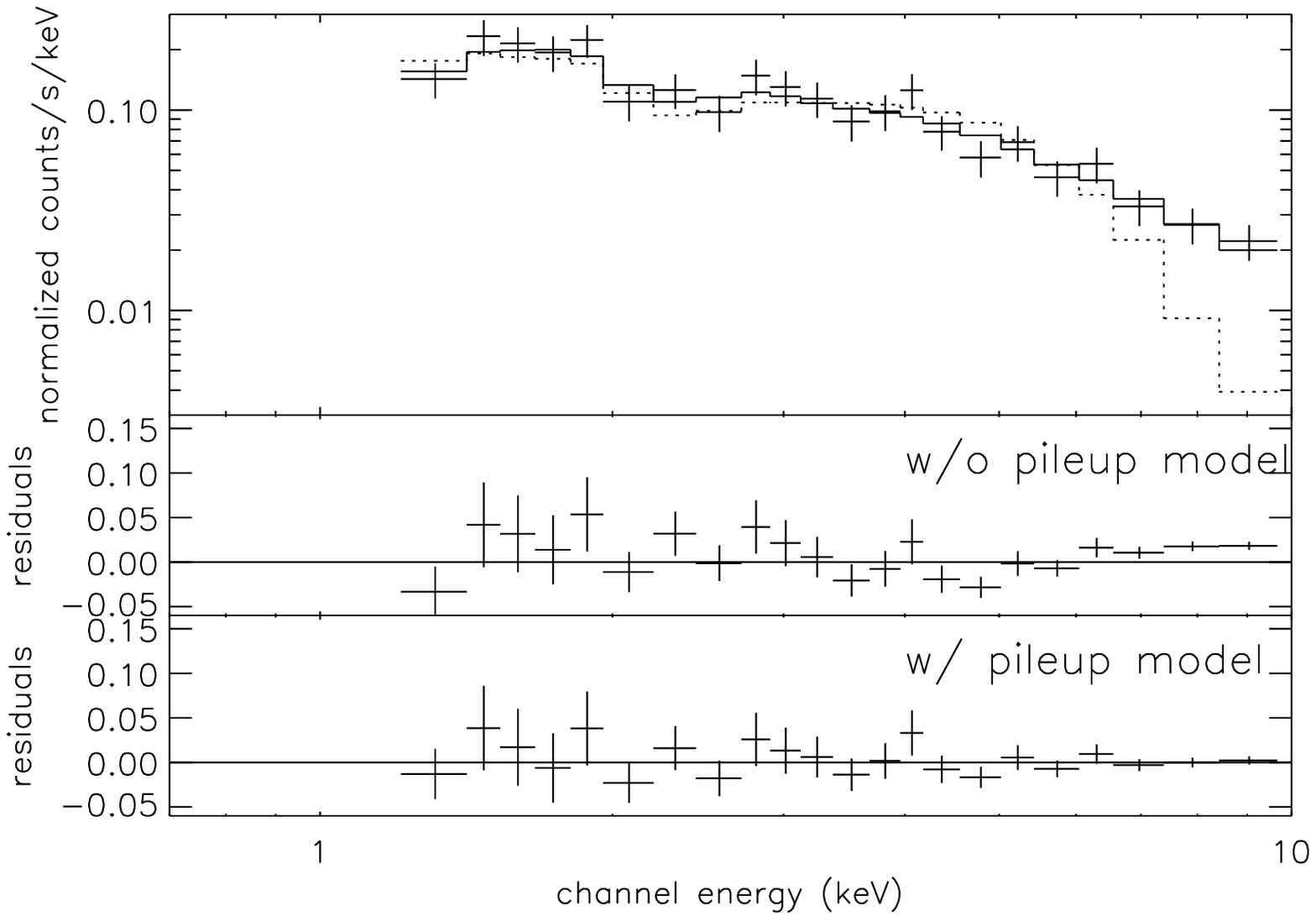}{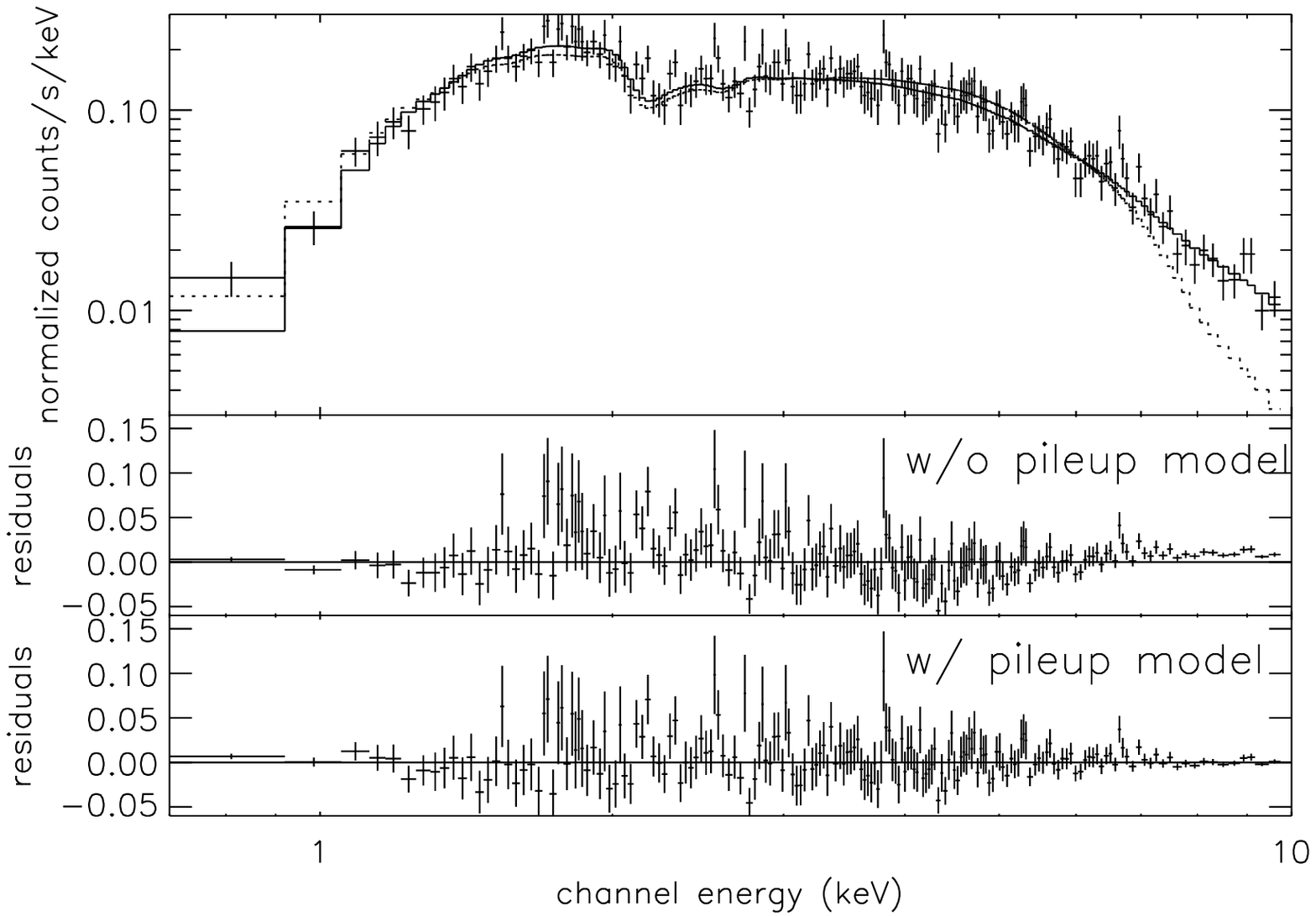}
\caption{Energy spectra extracted from TE mode images, fitted using an absorbed
power-law model (dotted line) and a piled-up absorbed power-law (solid line) for
2S 1711--339 (left) and  1E 1746.7--3224 (right).
\label{fig:pile-up}}
\end{figure}

Energy spectra were difficult to study for most of these sources because in all
but one of the observations where a source was detected, we had only imaging
(TE) mode data and pile-up was a problem. For these sources we used two
approaches to extract and fit the energy spectra: (1) direct extraction from the
TE mode data, fitting the data using the pile-up model and (2) extraction of 
trailed image spectra. To directly extract spectra and associated response files
for each source, we used a source region consisting of a 6\arcsec\ radius circle
and a background region consisting of an annulus with inner and outer radii of 
6\arcsec\ and 30\arcsec, respectively, in the CIAO tool {\it psextract}. These 
spectra were then fitted using XSPEC 11.2. For all sources, we fitted data for
energies from 0.3-10.0 keV using an absorbed power law model with pile-up
(phabs*powerlaw*pile-up). For most sources we found that the PSF fraction 
treated for pile-up (parameter 5) in the pile-up model, needed to be reduced 
from its default value of 0.95 to 0.85 to allow the grade morphing parameter 
(parameter 4) to not be pegged at 1.0 and to allow the model to fit the data at
higher energies. For some sources, the pile-up model had no effect on the 
spectral fits, even though the source was obviously piled-up; hence we used 
trailed images to characterize those spectra.  Results of our fits with the 
pile-up model are listed in Table~\ref{tab:spec}. Figure~\ref{fig:pile-up} shows
examples of spectra before and after the pile-up model is applied.

Trailed images were extracted using the CIAO tool {\em acisreadcorr}. During
each row transfer, the detector is exposed to the source for 40$\mu$s, but the
photons are recorded in the wrong Y position, resulting in a streak along the
x-axis. We calculated the total transfer streak exposure time using the
following expression
\begin{equation}
t^{\rm exposure}_{\rm trail} = t_1*(1024*f_{\rm sub}-dy)*t_{\rm exposure}/t_2
\end{equation}
where $t_1 = 40 \mu$s, the time to transfer charge from one row to the next;
1024 is the number of pixels in a row; $f_{\rm sub}=1/4$ is the subarray
fraction, $dy = 50$ is the number of pixels excluded near the source; 
$t_{\rm exposure}$ is the total exposure time in seconds; and $t_2 = 0.841040$ 
s, the time for one readout of the subarray. Spectra were extracted using the
CIAO tool {\em psextract} with the source region defined as two boxes along the 
y-axis, each 6 pixels wide in the x-direction. The two boxes included the
entire y-axis, except a region $\pm 25$ pixels from the y-coordinate of the 
piled-up source.  The EXPOSURE and BACKSCAL keywords in the spectra were
corrected for using the results of the CIAO tool {\em acisreadcorr} and the
calculation above. The resulting exposure times and total counts measured from
the source in each trailed image are listed in Table~\ref{tab:trspec}.

\begin{figure}
\plottwo{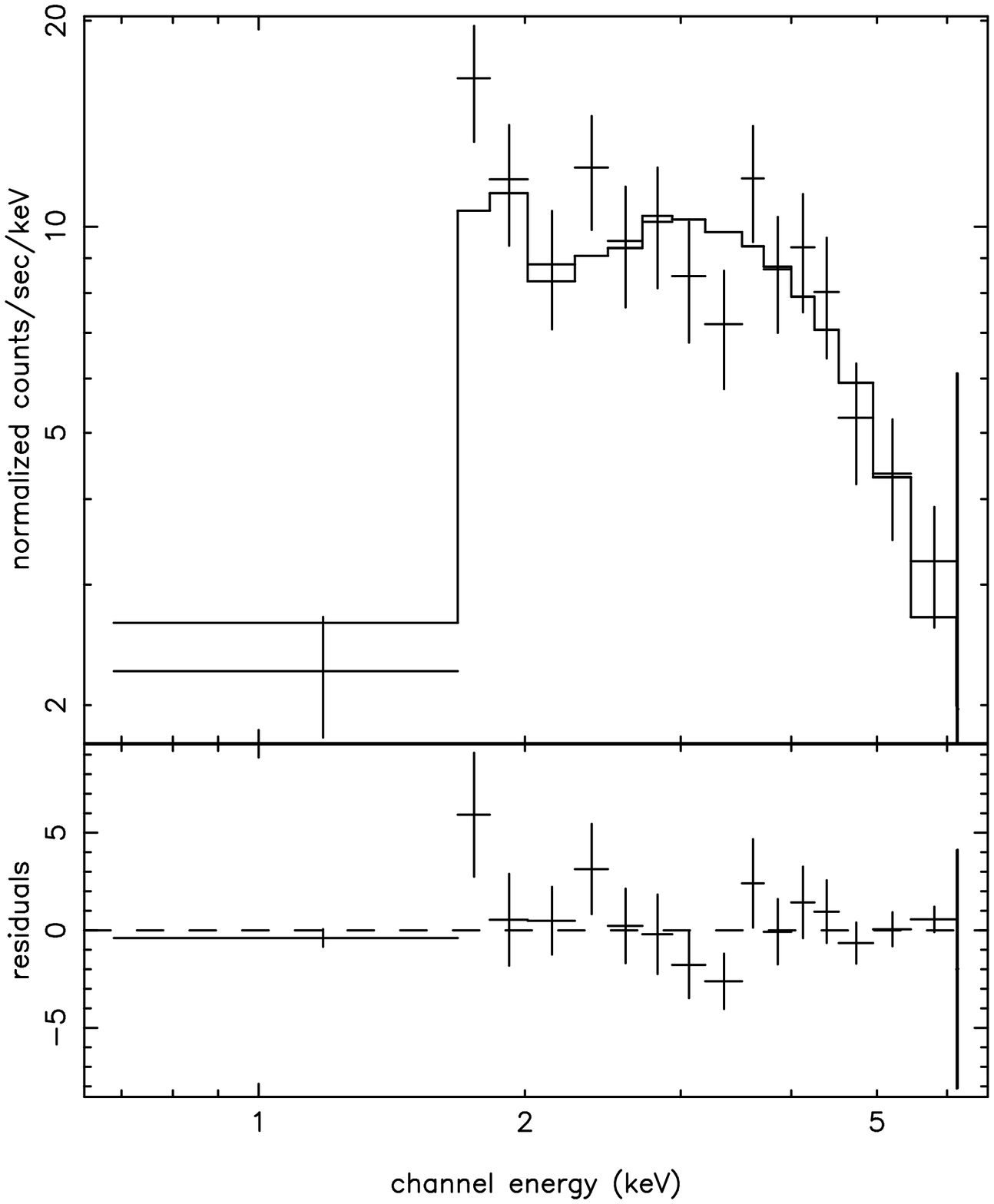}{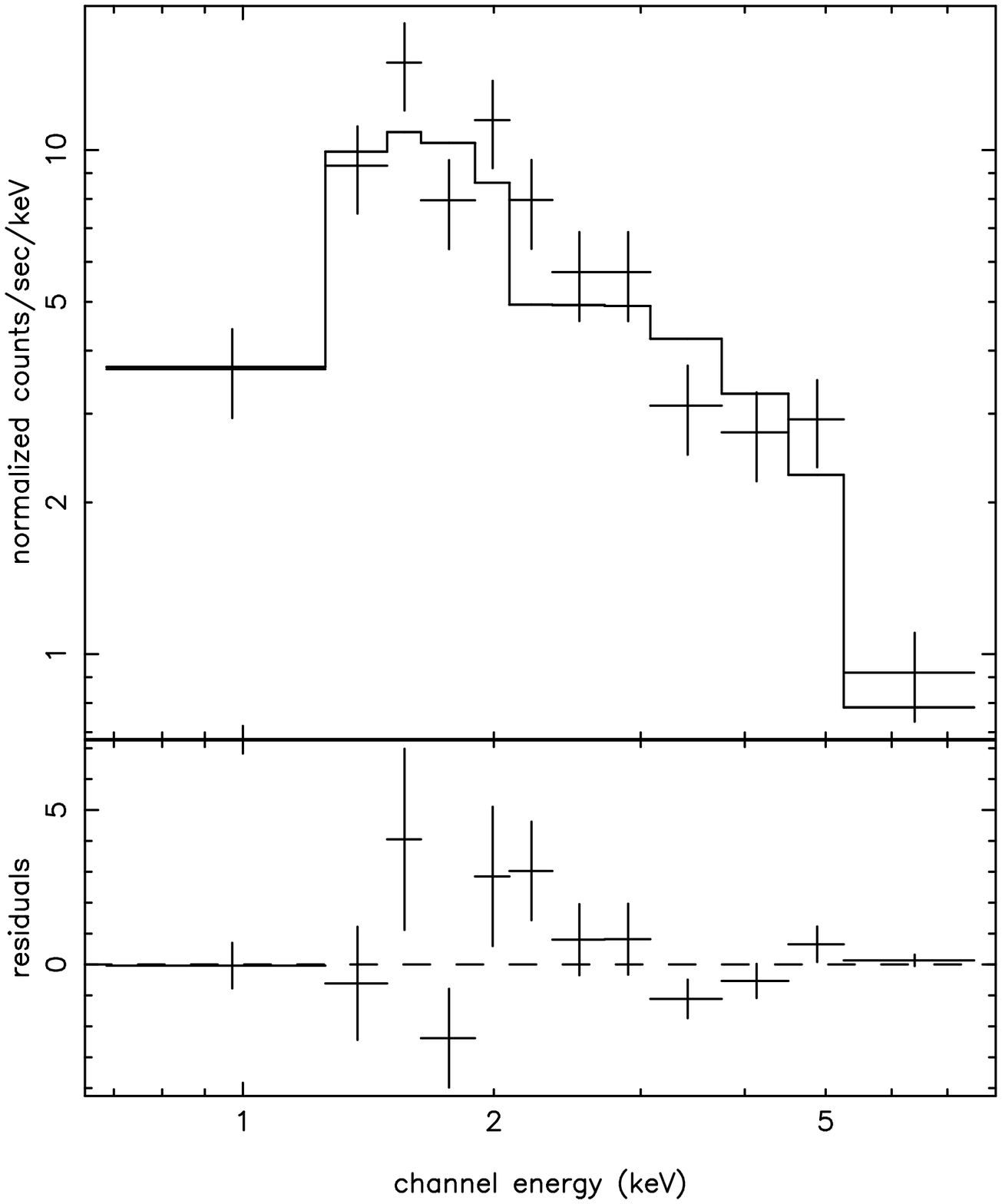}
\caption{Trailed image spectra for 4U 1708--408 (left) and 4U 1812--12 (right).
\label{fig:trailed}}
\end{figure}

For five of the observations, trailed image spectra could be extracted. We 
grouped the source spectra into bins containing 15 counts. Then these spectra 
were fitted in XSPEC 11.2 with an absorbed power-law model (phabs*powerlaw). For
two of the sources, 2S 1711--339 and 1E 1746.7--3224, very few counts remained 
in the trailed image spectra. The results of fitting trailed image spectra are 
listed in Table~\ref{tab:spec} and examples are shown in 
Figure~\ref{fig:trailed}.

\begin{figure}
\epsscale{0.9}
\plotone{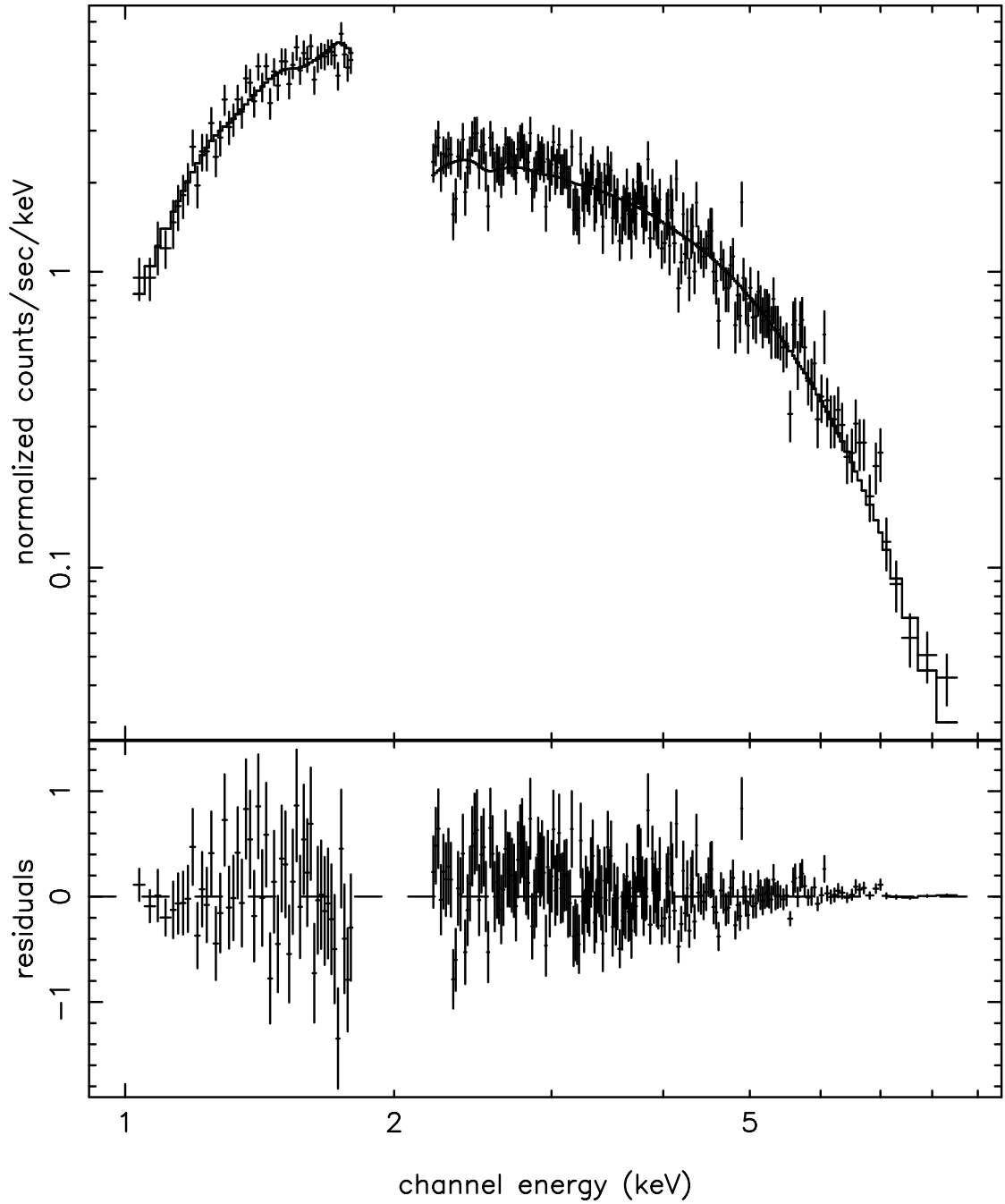}
\caption{Continuous Clocking mode energy spectrum for SLX 1735--269. This
observation fell on the S2 chip, a front illuminated CCD.
\label{fig:ccspec}}
\end{figure}

For one of the sources, SLX 1735--269, energy spectra were extracted from CC 
mode data taken on 2000 April 4. A source region consisting of a 6\arcsec\ 
radius circle and a background region consisting of an annulus with inner and 
outer radii of 6\arcsec\ and 30\arcsec, respectively, was used in the CIAO tool 
{\it psextract} to extract source and background spectra for the CC mode data 
and to create associated detector response files. We then grouped the data in 
the source spectrum so that each bin contained 25 counts. This observation fell
on the ACIS-S2 chip, a front illuminated chip, so the low-energy response was 
reduced. In XSPEC\footnote{http://heasarc.gsfc.nasa.gov/docs/software/lheasoft/xanadu/xspec/index.html}
11.2 \citep{Arnaud96}, we fitted data for energies of 1.0-1.8 keV and 2.2-9.0 
keV with an absorbed power-law model (phabs*powerlaw). Fit results are listed in
Table~\ref{tab:spec} and are shown in Figure~\ref{fig:ccspec}. 

The CC mode observations of SLX 1735--269 resulted in the largest number of 
source counts of any of our sources. Hence we used these observations to
investigate if more complicated spectral models are warranted. For example,
\citet{Migliari03} reported an iron line with energy $6.5 \pm 0.1$ keV, fixed
width 0.9 keV, and equivalent width 344 eV in addition to an absorbed power-law
in observations of 4U 1708--408. To search for evidence of a similar line in SLX
1735--269, we included a Gaussian with a fixed centroid at 6.5 keV and fixed 
width of 0.9 keV. Our fit resulted in a 3-sigma upper limit on the equivalent 
width of 975 eV, indicating that we had insufficient statistics to detect such a
line.

Next, since a soft thermal component has been observed in several LMXBs, we 
fitted a blackbody in addition to the absorbed power-law.  The fit gave
$\chi^2 = 278.2$ with 253 degrees of freedom. An F-test comparing this fit to
the absorbed power law had a value of 6.02 and a probability of $2.8 \times
10^{-3}$, suggesting that a softer component may be present. The 
resulting fit parameters were: $N_{\rm H} = (1.3 \pm 0.1) \times 10^{22}$ cm$^{-2}$, 
blackbody temperature $kT = 0.56 \pm 0.03$ keV, blackbody normalization 
$(R_{\rm bb}/d_{\rm 10\ kpc})^2 = 71 \pm 15$, power law index $= 1.4 \pm 0.2$, 
and power law normalization $= (2.5 \pm 1.1) \times 10^{-2}$ photons keV$^{-1}$
cm$^{-2}$ s$^{-1}$ at 1 keV. Since this observation with the largest number of
source counts resulted in only a
suggestion of a softer component, we concluded that we had insufficient
statistics to distinguish spectral models in the rest of our observations, which
had significantly fewer source counts.

\subsection{Timing}
Previous \rxte observations of SLX 1735--269 with 3-25 keV fluxes $\gtrsim 3.8
\times 10^{-10}$ \ergcms show band limited noise levels with fractional 
amplitudes of 24-28\% and break frequencies of 0.1-0.2 Hz \citep{Wijnands99b, 
Barret00, Belloni02} and a QPO near 0.9 Hz with a fractional rms of $\sim 5$\% and
FWHM = 0.3-0.8 Hz. \citet{Wijnands99b} also reported a very different power \
spectrum from ``low count rate" 1997 \rxte observations of SLX 1735--269 with 3-25 keV fluxes 
$\lesssim 2.8 \times 10^{-10}$ \ergcmsnospace, that had a break frequency of 
2.3 Hz and a fractional rms of 17\%. \citet{Barret03} fitted \sax power spectra
for 4U 1812--12 with a 4 Lorentzian model after \citet{Belloni02}. The 
low-frequency Lorentzian, fitting the low-frequency end of the band limited 
noise, had a fractional amplitude of 9.8\% rms and a break frequency of 0.15 Hz.
A low-frequency QPO was also detected with \sax at 0.73 Hz with a fractional 
amplitude of 3.9\% rms. 

We computed the sensitivity of our Chandra observations to features in the
power spectrum using
\begin{equation}
r = (2 N_{\sigma})^{1/2} \frac{(S+B)^{1/2}}{S} 
\left( \frac{\Delta \nu}{T} \right)^{1/4} \label{eqn:sens}
\end{equation}
where $r$ is the fractional root-mean-squared (rms) of the signal; $N_{\sigma}$
is the significance of the expected detection; $S+B$ is the total observed 
count rate; $S$ is the source count rate; $\Delta \nu$ is the full-width 
half-max (FWHM) of the signal, which is approximately equal to twice the break 
frequency obtained with a broken power law model for band limited noise (see 
Belloni, Psaltis, \& van der Klis 2002); and $T$ is the exposure time. For 
TE-mode observations of SLX 1735--269, 4U 1708--408, 2S 1711--339, 4U 1812--12,
and the 2000 Aug 30 observation of 1E 
1746.7--3224, $S+B$ = 0.2-0.7 cts s$^{-1}$, $S$ = 0.2-0.7 cts s$^{-1}$, and $T$
= 950-6800 s, resulting in 3-$\sigma$ sensitivities of 30-50\% rms for 
features with $\Delta \nu  = 0.3$ Hz. The 2002 July 16 observation of 1E 
1746.7--3224 was longer, with $T$ = 8500 s, and had slightly higher count rates
of $S+B =1.2$ cts s$^{-1}$ and $S = 1.2$ cts s$^{-1}$, resulting in a 3-$\sigma$
sensitivity of 23\% rms. Our \chandra TE mode observations were not
sensitive enough to detect broadband features at previously detected levels.
However, the CC mode observation of SLX 1735--269, with $S+B = 10.3$ cts
s$^{-1}$, $S = 10.2$ cts s$^{-1}$, and $T = 1400$ s, had a much better 
3-$\sigma$ sensitivity of 9.3\% rms for $\Delta \nu = 0.3$ Hz. Hence band
limited noise at levels observed in the brighter state of SLX 1735--269 
\citep{Wijnands99b,Barret00, Belloni02} should have been detectable in the CC 
mode observations; however, the observed QPO was not detectable. We estimated a 
3-$\sigma$ sensitivity of 18\% rms for $\Delta \nu = 4.6$ Hz, corresponding to
the fainter state observed by \citet{Wijnands99b}, for our CC mode observation 
of SLX 1735--269; hence we were unable to confirm this state with our 
\chandra observations. 

For the CC mode observation of SLX 1735--269, we created a light curve using 
CIAO tools\footnote{http://asc.harvard.edu/ciao/threads/aciscctoa/} that 
account for the fact that event times recorded in CC mode data were the times 
the events were read, not the times when the charge was deposited on the 
detector \citep{Zavlin00}. We generated a power spectrum for the energy range 
0.3-10 keV from the 2.85 ms CC mode light curve. No obvious variability was 
present in either the light curve or the power spectrum. Integrating from 
0.01-1 Hz, we estimated a noise strength of $\sim 10$\% rms for this 
observation; since this noise level was near our detection threshold, we cannot
easily confirm that this was entirely source related. However, the lack of
strong band limited noise at levels $> 20$\% rms, combined with our 1-10 keV 
flux measurement of $1.9 \times 10^{-10}$ \ergcms suggests that SLX 1735--269 
was most likely in a state similar to the ``lowest count rate" state reported 
in \citet{Wijnands99b}.

\subsection{UKIRT Images}

We obtained images of SLX 1735--269, 1E 1746.7--3224, and 4U 1812--12 using the
United Kingdom Infrared Telescope (UKIRT) to search for near-infrared
counterparts. For SLX 1735--269, the seeing was approximately 0.7\arcsec, the 
airmass was 1.46, the upper limit on the presence of a star in the \chandra 
error circle was $J>19.4$. The astrometric solution was performed using five stars
from the 2Mass in J catalog. The fit was good with an rms of 0.26\arcsec. The
astrometric accuracy of the 2Mass Catalogue is $<0.2\arcsec$. For 1E 1746.7--3224
the seeing was approximately 0.6\arcsec, the airmass was 1.64, and the upper 
limit for a source in the \chandra error circle was $J > 19.6$. The astrometric
solution was performed using eight stars from the 2Mass in J catalog. The fit was 
good with an rms of 0.23\arcsec. There was no 2Mass image available for
4U 1812--12; hence, the astrometric solution of the UKIRT image of the field
containing 4U 1812--12 (seeing  $\sim 0.6\arcsec$, airmass 1.18) was obtained by
identifying four USNO-A1.0 stars in an optical Digital Sky Survey (DSS) image with
a near infrared star. However, with an rms of 0.8\arcsec\ the astrometric
solution was poor. A star appears to be present near the error circle, but its 
J band magnitude is difficult to assess since we have no J band magnitudes for 
other stars in the field for comparison.

\section{Discussion}

Using \chandra we observed eight faint, little studied likely
LMXBs. Of the eight systems, we detected five in at least one observation: 4U
1708--408, 2S 1711--339, SLX 1735--269, 1E 1746.7--3224, and 4U 1812--12. The
\chandra observations reported in this paper resulted in precise locations for
all five of these objects, allowing for future deep optical and infrared
observations. UKIRT images of SLX 1735--269, 1E 1746.7--3224, and 4U 1812--12
did not reveal counterparts in the J band. Energy spectra for these five systems
were generally consistent with hard power laws (with photon index 1.5-2)
typical for faint LMXBs \citep[and references therein]{vanderKlis94,
vanParadijs94, Barret96}. Absorbed fluxes (1-10 keV) for the detected sources 
ranged from $(2.1-89) \times 10^{-11}$ \ergcmsnospace. The relatively narrow range of 
power law indices and observed fluxes suggest that all five of the detected systems
may be in a similar state. 

Energy spectra and fluxes for 4U 1708--408, SLX 1735--269, and 4U 1812--12 were
consistent with previous measurements in similar energy bands, supporting 
earlier findings that these systems are persistent. We observed SLX 1735--269 at
a 1-10 keV flux level of $2 \times 10^{-10}$ \ergcms in both observations. This
combined with the lack of detection of strong band limited noise in the CC mode
observation suggests that SLX 1735--269 was in a similar state to the ``low
count rate" state reported in \citet{Wijnands99b}. The energy spectrum and flux
measured using the pile-up model with \chandra on 2000 June 9 for 2S 1711--339 
were consistent with that measured with the \sax NFI on 2000 Feb 29 and with 
\sax WFC upper limits on 2000 March 22, suggesting that we were observing either
an extended tail of the bright 1998-1999 outburst observed with \sax and \rxte 
or a faint outburst. An estimate of the 2-6 keV flux reported in 
\citet{Cornelisse02} from the 2000 June 9 \chandra observation was a factor of 
$\gtrsim 10$ fainter than the flux we measured with \chandra for the same 
observation. We believe that \citet{Cornelisse02} did not properly account for 
the effects of pulse pile-up. In our second \chandra observation of 2S 1711--339,
in 2002 March, it had faded below detectability in CC mode. 1E 1746.7--3224, the
least studied of our five detected sources, and the only one of the five not to
have
previously shown X-ray bursts, was observed with \chandra at absorbed 1-10 keV 
fluxes of $2.1 \times 10^{-11}$ \ergcms and $(3.2-3.3) \times 10^{-11}$ \ergcms
in 2000 and 2002, respectively. In the \rosat all-sky survey
\citep{Voges99}, 1E 1746.7--3224 was measured at a count rate of 0.08 counts
s$^{-1}$ in the Position Sensitive Proportional Counter (PSPC). This corresponds to an absorbed 1-10 keV flux of
$(2-3) \times 10^{-11}$ \ergcmsnospace, if the spectral shape we measured with 
\chandra is assumed. These fluxes are a factor of $\sim 10$ higher than that 
observed in the \einstein galactic plane survey \citep{Hertz84}. The similarity
of the \chandra and \rosat fluxes suggests this source is persistent, and 
the \einstein measurement suggests that it, like most persistent sources, is
also variable. The three undetected systems, GRS 1736--297, KS 1739--304, and SLX 1746--331, had shown previous 
transient behavior. These non-detections were not due to positional errors. 
Upper limits on their fluxes were $(0.2-1.2) \times 10^{-13}$ \ergcmsnospace, two
to three orders of magnitude fainter than our faintest detection. If these 
sources were located at the galactic center, at a distance of 8 kpc, then these fluxes
would correspond to luminosities of $(2-9) \times 10^{32}$ ergs s$^{-1}$.

\acknowledgements
The United Kingdom Infrared Telescope is operated by the Joint Astronomy 
Centre on behalf of the U.K. Particle Physics and Astronomy Research
Council, and some of the data reported here were obtained as part of the UKIRT
Service Programme. CK and SP are party supported by SAO grants GO0-1054A and 
GO2-3046B. WHGL is grateful for support from NASA.

\begin{deluxetable}{lcccc}
\tabletypesize{\scriptsize}
\tablecaption{Summary of Observed X-ray Fluxes\tablenotemark{a}}
\tablewidth{0pt}
\tablehead{\colhead{Object} & \colhead{Date} & \colhead{Instrument} & \colhead{Energy Range} &
\colhead{Flux (\ergcms)}}
\startdata 
4U 1708--40 & 1996-1997 & \rxte ASM & 2-10 keV & $6.2 \times 10^{-10}$ \\
\nodata & mid-1998 & \rxte ASM & 2-10 keV & $1.0 \times 10^{-10}$ \\
\nodata & 1999 August & \sax & 2-10 keV & $1.2 \times 10^{-10}$\tablenotemark{b} \\
\nodata & 2000 May 15 & \chandra & 1-10 keV & $8.7 \times 10^{-10}$ \\
\nodata & 2000 Jun 18 & \rxte PCA & 2-10 keV & $7.4 \times 10^{-10}$\tablenotemark{b} \\
\nodata & mid-2001 & \rxte ASM & 2-10 keV & $8.3 \times 10^{-10}$ \\
\nodata & 2001 August & \sax & 2-10 keV & $8.9 \times 10^{-10}$\tablenotemark{b} \\
2S 1711--339 & 1998 July - 1999 May & \rxte ASM & 2-10 keV & $8.2 \times 10^{-10}$ \\
\nodata & 1998 July - 1999 May & \sax WFC & 2-28 keV & $6.3 \times 10^{-10}$ \\
\nodata & 2000 February 29 & \sax NFI & 2-6 keV & $2.4 \times 10^{-11}$ \\
\nodata & 2000 March 22 & \sax WFC & 2-28 keV & $\lesssim 7 \times 10^{-11}$ \\
\nodata & 2000 June 9 & \chandra & 1-10 keV & $4.4 \times 10^{-11}$ \\
\nodata & 2002 March 12 & \chandra & 1-10 keV & $\lesssim 1 \times 10^{-13}$ \\ 
SLX 1735--269 & 1996-2003 & \rxte ASM & 2-10 keV & $3.1 \times 10^{-10}$ \\
\nodata & 1997 February-May & \rxte PCA & 3-25 keV & 
 $(2.8-3.8) \times 10^{-10}$ \\
\nodata & 1997 October & \rxte PCA & 1-20 keV & $7.2 \times 10^{-10}$ \\
\nodata & 2000 April 4 & \chandra & 1-10 keV & $1.9 \times 10^{-10}$ \\
\nodata & 2000 May 23 & \chandra & 1-10 keV & $2.1 \times 10^{-10}$ \\
GRS 1736--297 & 1990 September-October & \granat ART-P & 3-12 keV & $6 \times
10^{-11}$ \\
\nodata & 2000 May 31 & \chandra & 1-10 keV & $\lesssim 8 \times 10^{-14}$ \\
KS 1739--304 & 1989 & \mir/KVANT & 2-30 keV & $2 \times 10^{-10}$ \\
\nodata & 1990 & \granat ART-P & 8-20 keV & $\lesssim 5 \times 10^{-11}$  \\
\nodata & 2002 May 5 & \chandra & 1-10 keV & $\lesssim 1 \times 10^{-13}$ \\
SLX 1746--331 & 1985 August & Spacelab-2 XRT & 2-10 keV & $6.35
\times 10^{-10}$\\
\nodata & 2000 June 9 & \chandra & 1-10 keV & $\lesssim 2 \times 10^{-14}$ \\
1E 1746.7--3224 & 1978-1981 & \einstein & 2-10 keV & $3 \times 10^{-12}$ \\
\nodata & 1985 August & Spacelab-2 XRT & 2-10 keV & $\lesssim 4 \times 
 10^{-11}$ \\
\nodata & 1990-1991 & \rosat PSPC & 1-10 keV & $(2-3) \times 10^{-11}$ \\
\nodata & 2000 August 30 & \chandra & 1-10 keV & $2.1 \times 10^{-11}$ \\
\nodata & 2002 July 15-16 & \chandra & 1-10 keV & $(3.2-3.3) \times 10^{-11}$ \\
4U 1812--12 & \nodata  & \oso, \arielv, \heao, \exosat & 2-10 keV & $4 \times
10^{-10}$ \\
\nodata & 1996-2003 & \rxte ASM & 2-10 keV & $3.7 \times 10^{-10}$ \\
\nodata & 2000 June 14 & \chandra & 1-10 keV & $4.4 \times 10^{-10}$ \\ 
\enddata 
\label{tab:fluxsum}
\tablenotetext{a}{This table is intended as a convenient summary of observed
fluxes discussed in the text and is not meant to be a complete record of all
observations.}
\tablenotetext{b}{Unabsorbed flux.}
\end{deluxetable}

 
\begin{deluxetable}{lclcccc}
\tabletypesize{\scriptsize}
\tablecaption{\chandra Observations}
\tablewidth{0pt}
\tablehead{\colhead{Target Object} & \colhead{Observation ID} & \colhead{Date} & 
\colhead{ACIS-S Mode} & \colhead{Time Resolution} & 
\colhead{Exposure Time (ks)} & \colhead{Detected}}
\startdata
4U 1708--40 & 661 & 2000 May 15 & TE & 0.841 s & 1.2 & Y \\
2S 1711--339 & 662 & 2000 Jun 9 & TE & 0.841 s &1.0 & Y\\
\nodata & 2695 & 2002 Mar 12 & CC & 2.85 ms & 4.7 & N\\
SLX 1735--269 & 664 & 2000 Apr 4 & CC\tablenotemark{a} & 2.85 ms & 1.4 & Y \\ 
\nodata & 663 & 2000 May 23 & TE & 0.841 s & 1.7 & Y \\
GRS 1736--297 & 665 & 2000 May 31 & TE & 0.841 s & 1.3 & N\\
\nodata & 666 & 2000 Jul 7 & CC & 2.85 ms & 12.2 & N \\
KS 1739--304 & 2696 & 2002 Apr 27 & CC & 2.85 ms & 3.4 & N\\
\nodata & 2697 & 2002 May 5 & TE & 3.24 s & 0.8 & N \\
SLX 1746--331 & 667 & 2000 Jun 9 & TE & 0.841 s & 1.7 & N\\
\nodata & 668 & 2000 Jul 24 & CC & 2.85 ms & 4.1 & N\\
1E 1746.7--3224 & 669 & 2000 Aug 30 & TE & 3.24 s & 6.8 & Y \\
\nodata & 2698 & 2002 Jul 15-16 & TE & 0.841 s & 8.5 & Y \\
4U 1812--12 & 670 & 2000 Jun 14 & TE & 3.24 s & 1.0 & Y\\
\enddata
\label{tab:obs}
\tablenotetext{a}{This observation fell on the S2 chip, rather than the standard
S3 chip.}
\end{deluxetable}

\begin{deluxetable}{lll}
\tabletypesize{\scriptsize}
\tablecaption{Source Locations (J2000).\tablenotemark{a}}
\tablewidth{0pt}
\tablehead{\colhead{Object} & \colhead{Right Ascension} & 
\colhead{Declination}}
\startdata
4U 1708--40 & $17^h12^m23.^s83$ & $-40\arcdeg50\arcmin34.0\arcsec$ \\
2S 1711--339 & $17^h14^m19.^s78$ & $-34\arcdeg02\arcmin47.3\arcsec$ \\
SLX 1735--269 & $17^h38^m17.^s12$ & $-26\arcdeg59\arcmin38.6\arcsec$ \\
1E 1746.7--3224\tablenotemark{b} & $17^h50^m3.^s90$ & $-32\arcdeg25\arcmin50.4\arcsec$ \\
\nodata\tablenotemark{c}         & $17^h50^m3.^s95$ & $-32\arcdeg25\arcmin50.1\arcsec$ \\ 
4U 1812--12 & $18^h15^m06.^s18$ & $-12\arcdeg05\arcmin47.1\arcsec$ \\
\enddata
\tablenotetext{a}{90\% confidence error radius $\simeq 0.6$\arcsec\ on axis.}
\tablenotetext{b}{2000 Aug 30 observation.}
\tablenotetext{c}{2002 Jul 15-16 observation.}
\label{tab:loc}
\end{deluxetable}

\begin{deluxetable}{lllllllc}
\tabletypesize{\scriptsize}
\tablecaption{Spectral Fitting Results\tablenotemark{a}}
\tablewidth{0pt}
\tablehead{\colhead{Object} & \colhead{Date} & \colhead{Type\tablenotemark{b}} &
 \colhead{N$_{\rm H}$} & \colhead{Index} &
 \colhead{$\alpha$\tablenotemark{c}} & \colhead{$\chi^2/$dof} & 
 \colhead{Flux (1-10 keV)} \\
 \colhead{} & \colhead{} & \colhead{} & \colhead{($10^{22}$ cm$^{-2}$)} &
 \colhead{} & \colhead{} & \colhead{} & \colhead{$10^{-11}$ \ergcms}}
\startdata
4U 1708--40 & 2000 May 15 & Tr & $3.3\pm0.5$ & $1.9 \pm 0.3$ & \nodata & $14.2/15$
& 87 \\
2S 1711--339 & 2000 Jun 9 & TE & $1.5\pm0.3$ & $1.9 \pm 0.2$ & $0.64 \pm 0.02$ & 
$11.8/18$ & 4.4 \\
\nodata  &  \nodata& Tr & $0.9\pm0.4$ & $1.9 \pm 0.7$ & \nodata         & 
$2.8/2$ & 11  \\  
\nodata  & 2002 Mar 12 & CC & 1.4\tablenotemark{d} & 1.9\tablenotemark{d} &
\nodata & \nodata & $\lesssim 0.01$\tablenotemark{e} \\    
SLX 1735--269 & 2000 Apr 4 & CC & $1.70 \pm 0.05$ & $2.07 \pm 0.04$ & \nodata & 
$291.4/255$ & 19 \\
\nodata & 2000 May 23 & Tr & $1.8 \pm 0.5$ & $2.2 \pm 0.6$ & \nodata & $13.1/10$
& 21 \\
GRS 1736--297 & 2000 May 31 & TE & 1.1\tablenotemark{d} & 1.8\tablenotemark{d}
& \nodata & \nodata & $\lesssim 0.008$\tablenotemark{e} \\
KS 1739--304 & 2002 May 5 & TE & 1.4\tablenotemark{d} & 1.5\tablenotemark{d} 
& \nodata & \nodata & $\lesssim 0.01$\tablenotemark{e} \\
SLX 1746--331 & 2000 Jun 9  & TE & 0.4\tablenotemark{d} & 1.5\tablenotemark{f} &
\nodata & \nodata & $\lesssim 0.002$\tablenotemark{e} \\
1E 1746.7--3224 & 2000 Aug 30 & TE & $2.0 \pm 0.1$ & $1.54 \pm 0.06$ & 
$0.96 \pm 0.02$ & $40.2/44$ & 2.1 \\
\nodata & 2002 Jul 15-16 & TE & $1.5 \pm 0.1$ & $1.1 \pm 0.1$ & $0.51 \pm 0.06$ & 
 $186.1/183$ & 3.2 \\
\nodata & \nodata     & Tr & $1.3 \pm 0.7$ & $1.2 \pm 0.7$ & \nodata         &
 $0.1/1$    & 3.3 \\
4U 1812--12 & 2000 Jun 14 & Tr & $1.1 \pm 0.2$ & $1.5 \pm 0.3$ & \nodata & $16.4/9$ & 
 44 \\
\enddata
\tablenotetext{a}{Errors on spectral fitting parameters are 1$\sigma$.}
\tablenotetext{b}{Tr = Trailed image, TE = Timed exposure mode (with pile-up
model), CC = Continuous clocking mode}
\tablenotetext{c}{Pile-up parameter.}
\tablenotetext{d}{Assumed spectral parameter.}
\tablenotetext{e}{99\% confidence upper limit from WebPimms.}
\tablenotetext{f}{Thermal Bremsstrahlung spectrum with kT = 1.5 keV assumed.}
\label{tab:spec}
\end{deluxetable}

\begin{deluxetable}{llcc}
\tabletypesize{\scriptsize}
\tablecaption{Trailed Image Spectra Exposure and Source Counts}
\tablewidth{0pt}
\tablehead{ \colhead{Object} & \colhead{Date} & \colhead{Exposure (s)} &
\colhead{Total Source Counts}}
\startdata
4U 1708--40 & 2000 May 15 & 11.2 & 412 \\
2S 1711--339 & 2000 Jun 9 & 8.84 & 86 \\
SLX 1735--269 & 2000 May 23 & 16.0 & 228 \\
1E 1746.7--3224 & 2002 Jul 15-16 & 79.4 & 117 \\
4U 1812--12 & 2000 Jun 14 & 11.9 & 312 \\
\enddata
\label{tab:trspec}
\end{deluxetable}

\onecolumn

\end{document}